# Modeling magnetic-field-induced domain wall propagation in modulated-diameter cylindrical nanowires

J. A. Fernandez-Roldan[1], A. De Riz[2], B. Trapp[3], C. Thirion[3], J.-C. Toussaint[3], O. Fruchart[2] and D. Gusakova[2,*]

[1]*Institute of Materials Science of Madrid, CSIC, 28049 Madrid, Spain*

[2]*Univ. Grenoble Alpes, CNRS, CEA, Grenoble INP\*\*, INAC-Spintec, F-38000 Grenoble, France*

[3]*Univ. Grenoble Alpes, CNRS, Institut NEEL, F-38000 Grenoble, France*

**ABSTRACT**

Domain wall propagation in modulated-diameter cylindrical nanowires is a key phenomenon to be studied with a view to designing three-dimensional magnetic memory devices. This paper presents a theoretical study of transverse domain wall behavior under the influence of a magnetic field within a cylindrical nanowire with diameter modulations. In particular, domain wall pinning close to the diameter modulation was quantified, both numerically, using finite element micromagnetic simulations, and analytically. Qualitative analytical model for gently sloping modulations resulted in a simple scaling law which may be useful to guide nanowire design when analyzing experiments. It shows that the domain wall depinning field value is proportional to the modulation slope.

*Corresponding author: daria.gusakova@cea.fr

\*\*Institute of Engineering Univ. Grenoble Alpes

# I. INTRODUCTION

Intensive study is devoted to circular cross-section magnetic nanowires due to their unique fundamental properties and their potential applications in a number of advanced technologies such as data storage, sensing and biomagnetics [1-4]. Electrochemical growth of cylindrical nanowires within self-organized porous templates [5, 6] is continuously progressing, and is nearing compatibility with design of a three-dimensional race-track memory [7]. The information stored in this type of solid-state device would be encoded by magnetic domains separated by magnetic domain walls. Domain wall motion in wires may be achieved by applying an external magnetic field [5, 8, 9], spin-polarized current [10], spin waves or localized temperature gradients [11]. In addition, in a real device, the domain wall position must be precisely controlled. Control can be achieved by designing well-defined pinning centers, for example, by changing the composition of the wire [12] or by introducing geometrical irregularities during the fabrication process [5, 8, 14-17]. In the latter case, modulated diameter of the nanowire synthesized by electrodeposition can be used to control domain wall position by locally reducing its magnetostatic and exchange energy in the smaller cross-sectional parts [18]. Thus, diameter modulation plays the role of a potential barrier which implies that some threshold driving force must be applied to overcome the barrier.

Few studies have theoretically addressed domain wall motion, induced either by applied field or current, within modulated-diameter nanowires [19-23]. Moreover, no theoretical model quantifying the threshold driving force as a function of geometric parameters has yet been reported, although experimental systems are now available. Thus, a general scaling law describing the domain wall depinning phenomenon would be useful to assist experimental system design and analysis of experimental results.

In this paper, we combine both approaches: quantitative micromagnetic description of the domain wall's behavior and development of a simplified qualitative analytical model which provides a simple general scaling law and relates geometric parameters to the domain wall depinning field. We focus on the case of wall motion under the conservative driving force produced by a magnetic field applied along the wire's axis. In this case, the magnetic field contribution can be treated in the framework of a potential energy well. However, our approach is quite general and may be extended further to study the effect of other driving forces such as spin-polarized current.

## II. MICROMAGNETIC SIMULATIONS

To understand how geometry affects domain wall propagation in a modulated-diameter cylindrical nanowire when a magnetic field is applied, we used our home-made finite element software *feeLLGood* [22, 24]. The non-regular finite element mesh of *feeLLGood* accurately describes the given geometry. The Landau-Lifshitz-Gilbert equation was solved for magnetization dynamics [25] by a micromagnetic approach

$$\frac{\partial \mathbf{m}}{\partial t} = -\gamma_0 (\mathbf{m} \times \mathbf{H}_{eff}) + \alpha \left( \mathbf{m} \times \frac{\partial \mathbf{m}}{\partial t} \right), \quad (1)$$

where $\mathbf{m}=\mathbf{M}/M_S$ is the unit vector in the direction of the local magnetization, and $M_S$ is the spontaneous magnetization. The following notations were used: $\gamma_0 > 0$ for the gyromagnetic ratio, $H_{eff}$ for the effective field (including magnetostatic, exchange and applied field contributions), $\alpha$ for the Gilbert damping factor.

Cylindrical nanowires with a single smooth diameter modulation connecting two straight parts were considered. The smaller diameter was $2R_1$ and the larger was $2R_2$, as illustrated in Figure 1(a). The circle-based profile with $R_{mod} = ((R_2 - R_1)^2 + \lambda^2)/(4(R_2 - R_1))$ sketched in Figure 1(b) suitably describes experimental cases [5]. The wire axis was taken as the $z$ direction. The modulation of the length $\lambda$ was centered at $z=0$ and $L$ is the total length of the wire.

Magnetic charges were numerically removed from the extremities to imitate an infinitely long wire and to prevent nucleation of magnetization reversal; and Brown (Neumann) conditions were used for magnetization components. The micromagnetic parameters used were those of a permalloy material, with $M_S=800 \cdot 10^3$ A/m, $\gamma_0=2.21 \cdot 10^5$ m/(A·s), $\alpha=0.05$ and exchange constant $A_{ex}=1 \cdot 10^{-11}$ J/m. The system's geometry was discretized into tetrahedrons with dimensions not exceeding 2 nm. The initial magnetic configuration corresponded to a relaxed transverse-like tail-to-tail wall [Fig.1(c)].

The domain wall position and width were estimated by fitting the longitudinal magnetization component with the tanh-profile similar to the standard Bloch wall model [26]. The profile fitted was $m_z = tanh\big((z - z_0)/\Delta_{DW}\big)$, where $z_0$ is the position of the center of the domain wall, and $\Delta_{DW}$ is the domain wall parameter. For simplicity, this parameter is called the domain wall width hereafter. In contrast to the widely-used "dynamic" definition of the domain wall width proposed by Thiele [27, 28], our formula does not impose constraints on the geometry or on the domain wall dynamics. Indeed, Thiele's definition of domain wall width assumes that the stationary domain wall moves along the invariable cross-section wire without changing its profile. These conditions do not hold in our case as the variable cross-section affects both the domain wall profile and its width. Although the profile corresponds to the one-dimensional Bloch model [25] rather than to fully three-dimensional magnetization distribution, it fits the transverse-like domain wall width well enough for moderate-diameter wires. Consequently, we restricted our calculations to moderate diameters. Larger diameters (≥40 nm) may favor Bloch Point domain walls with significantly different dynamic properties [29, 30].

## A. Straight wire limit

In this section we show that our numerical implementation is consistent with existing results for straight wires. Similar to what is reported by Jamet et al. [30] the domain wall width estimated from a tanh-profile fit increases with the wire's radius, as illustrated by the open circles in Figure 2(b). This curve tends to the Bloch wall parameter value $\Delta_{Bloch}= \sqrt{4A_{ex}/\mu_0 M_s^2}$ at $R=0$ [25, 31]. No significant difference was observed between tanh-fit values obtained in the absence or in the presence of an applied field. This result contrasts with those reported for rectangular strips, where the domain wall width is strongly affected by the magnetic field applied due to deformation of its structure [32, 33]. This crucial difference underlines the leading role of dipolar energy for cylindrical wires.

Under the applied magnetic field, forward domain wall motion at speed $v$ is accompanied by its azimuthal rotation at frequency $f$ around the axis of the wire [29], as illustrated in Figure 2(a). In the one-dimensional model we expect [34]

$$v = \frac{\alpha \gamma_0 \Delta_{DW}}{1+\alpha^2} H_{app} \qquad (2)$$

and

$$f = \frac{\gamma_0}{2\pi(1+\alpha^2)} H_{app}. \qquad (3)$$

In our simulations, forward domain wall velocity increased with the applied field in a linear fashion, and, as expected, the slope depended on the wire's radius. Figure 2(c) superposes simulated points on the solid lines calculated using Eq. (2) and tanh-profile estimations of the domain wall width. For wires with a small radius (~2.5 nm), simulations perfectly agreed with the analytical expression. Thus, we can conclude that domain wall behavior is well reproduced by the one-dimensional model. We attribute the moderate discrepancy between simulated and analytical values for larger radii (>5 nm) to the deviation of the real wall profile from the perfect tanh-profile shape. This discrepancy is also observable in Figure 2(b) where red stars depict the domain wall widths calculated from the curve representing simulated velocities plotted against applied field dependencies.

In contrast to the slight differences for linear domain wall velocity, perfect matching between Eq.(3) and numerical values was found for the domain wall rotation frequency [Figure 2(d)]. In this case, as expected, the domain wall's precessional frequency was independent of its width.

**B. Nearing and overcoming the modulation**

The presence of diameter modulation results in variation of the internal energy of the system depending on the position of the domain wall. To quantify this phenomenon, in Figure 3(a) we plotted the internal energy as a function of the coordinates obtained with domain walls drifting freely from the broader part toward the thinner part of the wire in the absence of applied field. As expected, the energy of the domain wall was minimized in the thinner part of the wire, because it is proportional to the wire's diameter at any given position, as detailed below in Section III. The highest energy difference corresponded to the highest $R_2/R_1$ ratio. The free drift was accompanied by non-monotonic modification of wall width [Figure 3(b)]. Similar behavior was reported elsewhere [20, 22]. We hypothesize that the reduced domain wall width to the left of the modulation and its increase to the right of the modulation can be attributed to the spatial modification of the magnetic charge when entering or leaving the modulation. The domain wall width is adapted in order to ensure the total magnetic charge conservation in the volume, schematized in Figure 3(c). Far from the modulation, both the energy and the width of the domain wall recover the values observed in a straight wire.

Next, we prepared the domain wall in the narrow section of the wire and drove it toward the larger section by applying a magnetic field. Figure 4(a) shows the domain wall position as a function of time; the changes to internal energy and domain wall width depending on its position are presented in Figures 4(b) and 4(c). Below a critical field value, the domain wall slows down when approaching the modulation, partially enters the modulation and finally comes to a halt.

Above a critical field value, the domain wall slows down but never comes to a complete standstill, it can thus overcome the energy barrier and continue traveling toward the right. Far to the right of the modulation, the domain wall moves at a constant rate according to Eq.(2). Its velocity is greater than in the narrow part of the wire. Near to the modulation, the domain wall velocity has non-trivial behavior which is not reflected by the simplified formula. To estimate the final position of the domain wall and the critical field intensity required to unpin it, we developed the analytical approach presented in the following section.

### III. ANALYTICAL APPROACH

In this section, we present a qualitative analytical model describing various features of the domain wall in the modulated-diameter wire. In particular, we establish an analytical scaling law which relates the critical applied field value to the geometry of the modulation. To determine this qualitative expression, we considered the domain wall's energy, based on the magnetostatic, exchange and applied field contributions. We also took the interaction between the domain wall and the magnetic charges induced by the modulation into account. For a circular cross-section wire [31] the domain wall energy reads:

$$E = E_0 + E_{mod},$$

$$E = \frac{\mu_0}{2}\int_V \boldsymbol{H}_d^2(\boldsymbol{r})dV + \int_V A_{ex}(\nabla \boldsymbol{m}(\boldsymbol{r}))^2 \, dV - \mu_0 M_s \int_V \boldsymbol{m}(\boldsymbol{r})\cdot \boldsymbol{H}_{app} \, dV + E_{mod}. \quad (4)$$

Below, we introduce the approximations that can be used to estimate each term.

For the dipolar energy, we considered that the magnetic charge $q_{wall} = -2\pi R^2 M_s$ [35] carried by the tail-to-tail wall was uniformly distributed over a sphere of radius $R$ with magnetic charge density $\rho = 3q_{wall}/4\pi R^3$ [Fig.5(a)]. The real distribution of the magnetic charge is much more complex [36, 37]. Nevertheless, our approximation is justified by the quasi linear relationship between the wall width and the wire radius in the dimensions considered [Fig.2(b)]; it leads to a compact analytical expression for the different energy terms and gives a reasonable order of magnitude. By analogy with electrostatics, the magnetostatic field $H_d$ is $H_d(r > R) = q_{wall}/4\pi r^2$ outside the sphere and $H_d(r < R) = q_{wall}r/4\pi R^3$ inside the sphere. Integration over the whole space reduces the magnetostatic energy contribution to $3\pi\mu_0 M_s^2 R^3/5$.

The exchange energy contribution in Eq.(4) can be estimated by applying the one-dimensional spin chain model [25] with slowly varying magnetization. In this case $(\nabla \mathbf{m}(r))^2 \approx (\pi/2R)^2$ and the integration of the second term in Eq.(4) over the sphere volume an exchange energy contribution equals to $A_{ex}\pi^3 R/3$.

To estimate the Zeeman energy contribution, we neglected the inner structure of the domain wall and considered the Zeeman energy of two adjacent uniformly magnetized domains located at the domain wall's center position, *z*. With this assumption, the energy contribution would be equal to $-2\mu_0 M_s H_{app}\pi R^2 L$ in the case of a straight wire. For the modulated-diameter wire, it would be equal to $-2\mu_0 M_s H_{app}\pi \int_{-\frac{L}{2}}^{z} R^2(\zeta)d\zeta + cts$. The domain wall energy without interaction then becomes

$$E_0(z) = \frac{3\pi}{5}\mu_0 M_s^2 R^3(z) + \frac{A_{ex}}{3}\pi^3 R(z) - 2\mu_0 M_s H_{app}\pi \int_{-L/2}^{z} R^2(\zeta)d\zeta + cts. \quad (5)$$

The domain wall energy estimated is similar to the energy values extracted from the simulations [Fig.5(b)] and thus is suitable to qualitatively describe domain wall behavior.

The domain wall interacts with magnetic charges induced by the change in diameter, as illustrated in Figure 5(a). This phenomenon gives rise to a supplementary energy term $E_{mod}$, the derivative of which is related to the magnetic field generated by the magnetic charges of the modulation: $\partial E_{mod}/\partial z = -\mu_0 q_{wall} H_{mod}(z)$. This term is unlikely to have any analytical expression with the geometry studied here. The field distribution $H_{mod}(z)$ can be analytically calculated by making some assumptions. For simplicity, we assume that only surface charges are induced by the magnetization parallel to the *z* axis inside the modulation. As a result, the field generated by these surface charges at any point along the central axis [38] may be calculated as

$$H_{mod}(z) = \frac{M_s}{2} \int_{charged\ surf.} \frac{sign(z-\zeta)(z-\zeta)R(\zeta)R\prime(\zeta)d\zeta}{\left(\sqrt{R^2(z)+(z-\zeta)^2}\right)^3}. \qquad (6)$$

The distribution of the $H_{mod}(z)$ field for different geometries is shown in Figure 5(d). This field opposes the magnetic field applied, as illustrated in Figure 5(a), and amplifies the domain wall's repulsion from the modulation. It is probable that the surface charge simplification overestimates the amplitude of $H_{mod}(z)$. Nevertheless, this approach provides an upper limit for $H_{mod}(z)$.

For the analytical calculation, the circle-based wire profile used in the numerical simulations may be approximately replaced by the following analytical differentiable function: $R(z) = \left(R_1 + R_2 + (R_2 - R_1)tanh(4z/\lambda)\right)/2$. This type of approximation is appropriate for use with the gently sloping modulations studied in this paper [39]. To illustrate this case, the domain wall energy $E_0$ is plotted as a function of the domain wall's position for different applied field values in Fig.5(c). Extending this figure by including the $E_{mod}$ term, if its analytical expression were available, would have no qualitative impact on the illustration, and would only result in shifts in the energy minima and maxima.

The position of the energy minimum, $\partial E(z)/\partial z = 0$, corresponds to the position of a pinned domain wall. Moreover, the domain wall depinning condition, at a given critical applied field value $H_{crit}$, can be defined as the convergence of two extrema at the same point – the inflection point. The equation linking the final domain wall position *z* to the applied magnetic field $H_{app}$ reads

$$(1 - tanh(4z/\lambda)^2)\left(1 + \frac{10l_{ex}^2\pi^2}{27\left(R_1 + R_2 + (R_2 - R_1)tanh(4\,z/\lambda)\right)^2}\right)\frac{(R_2 - R_1)}{\lambda} +$$

$$+\frac{5}{9M_s}\left(H_{app}+H_{mod}(z)\right)=0. \tag{7}$$

At this point, we introduced the exchange length $l_{ex}^2 = 2A_{ex}/\mu_0 M_s^2$. The numerical solution of this equation is illustrated by the solid black line in Figure 5(e). Equation 7 can be considerably simplified for gently sloping modulations, when $(R_2 - R_1)/(R_2 + R_1) \ll 1$ and $|H_{mod}| \ll |H_{app}|$. In this case, the final position of the domain wall can be determined analytically:

$$z_f = -\lambda \tanh^{-1}\left(\sqrt{1+kH_{app}}\right)/4, \text{ where} \tag{8}$$

$$k = \frac{5\lambda}{9M_s(R_2-R_1)}\left(1+\frac{10 l_{ex}^2 \pi^2}{27(R_1+R_2)^2}\right).$$

This analytical dependence is illustrated by the dashed red line in Figure 5(e). The corresponding critical field is given by

$$H_{crit} = \frac{9M_s(R_2-R_1)}{5\lambda}\left(1+\frac{10 l_{ex}^2 \pi^2}{27(R_1+R_2)^2}\right). \tag{9}$$

This approximated relation gives the lower limit for $H_{crit}$. Magnetostatic domain wall repulsion from a modulation due to surface poles, when not negligible, shifts $H_{crit}$ towards higher values. Nevertheless, the analytical formula (9) provides a good estimation of the order of magnitude of $H_{crit}$ and the relation between $H_{crit}$ and geometric parameters. According to Eq.(9), the critical field is simply proportional to the slope of the modulation $(R_2 - R_1)/\lambda$ with a negligibly small exchange correction for large diameters. Thus, $H_{crit}$ is almost a linear function of $R_2$ in Figure 6(a) and naturally equals zero in the case of a straight wire, where $R_1 = R_2$. The inverse proportionality between $H_{crit}$ and the modulation length $\lambda$ is illustrated in Figure 6(b). This relation verifies the moderate values of $H_{crit}$ for gently sloping modulations.

## IV. ANALYTICS vs. MICROMAGNETICS

Figure 7 compares the critical field to the modulation parameters obtained from micromagnetic simulations and those predicted by Eqs. (7) and (9). This comparison reveals qualitatively similar tendencies. Moreover, small $R_2/R_1$ ratios and long $\lambda$, corresponding to gently sloping modulations, ensure the best fit between the simulations and the analytical expression. These parameters also provide the best fit for expression describing the final domain wall position as a function of the applied field [Figure 8]. Indeed, of the three simulations tested, that with the gentlest sloping modulation [Fig.8(b)] showed the best agreement with Eqs.(7) and (8). In this case, the slope of the modulation satisfies the $(R_2 - R_1) \ll \lambda$ condition and $|H_{mod}| \ll |H_{app}|$, which reduces the magnetostatic contribution of the magnetic pole density stored on the modulation surface [40] and ensures better consistency between the tanh-based and circle-based profiles [39]. Under these conditions, the analytic formulae Eqs. (8) and (9) become a valuable tool for experimenters. These formulae can be helpful when determining the wire diameter and modulation sizes during the fabrication process to attain the most favorable conditions for domain wall depinning and propagation.

The comparison between analytical expressions and simulated values is less accurate for very abrupt geometries. In these cases the deviation between the two approaches becomes pronounced due to several assumptions used in the model. One of the approximations relates to magnetic charge distribution inside the domain wall. The approximation of a uniformly charged sphere turns out to be too rough and is not well reproduced in micromagnetic simulations [Figure 9]. Although the magnetic charge is concentrated in the confined region close to the center of the domain wall, its spatial distribution is far from constant within this area. The magnetic charge isovalues for the fully three-dimensional magnetic distribution have a non-trivial shape, with a maximum at the center of the domain wall. Another approximation showing limitations relates to the estimated exchange energy. The hypothesis of slow linear variation of the magnetization angle used here, by analogy with the one-dimensional spin chain model, nears its limit with abrupt modulations. In addition, the approximation of the uniformly charged sphere may be too rough for radii significantly larger than the exchange length. To deal with this type of case, the domain wall width should be scaled as the square of the radius [30]. Despite these limitations, our simplified analytical approach gave a very reasonable estimation of the behavior of the critical depinning field in response to the modulation parameters. The cases for which the assumptions used in our model are too drastic should be covered by micromagnetic simulations.

Here, our analytical model was focused on a transverse-like domain wall. Typically, dimensions of a few tens of nanometers, with a maximum diameter value between 30 and 50 nm, should ensure transverse-like domain wall stability for most common ferromagnetic materials [41] and thus allow application of our simplified model. Larger diameters favor the so-called Bloch Point Wall configuration. Although the magnetic charges and energy distribution for both types of walls are very similar for large diameters [30], an open question is whether the linear relationship between the depinning field $H_{crit}$ and the modulation slope $(R_2 - R_1)/\lambda$ holds for magnetic textures that are different from transverse-like walls. The result of the corresponding experiments would provide clues for further numerical and analytical analysis.

## V.   CONCLUSION

This paper presents a theoretical study of transverse domain wall behavior under the influence of an applied magnetic field in a circular cross-section nanowire. In particular, we investigated the domain wall pinning phenomenon close to the wire diameter modulation. We proposed both a quantitative micromagnetic description of the domain wall behavior using our finite element micromagnetic software, and a simplified analytical model which relates geometric parameters to the critical magnetic field needed to unpin the domain wall. Numerical simulations and analytical predictions based on an approximation of uniformly magnetized magnetic spheres agreed best in the case of gently sloping modulations. To provide a more accurate analytical description of domain wall behavior close to the modulation may require some correction of the model or the inclusion of additional energy terms. These terms may concern such phenomena as, for example, magnetocrystalline anisotropy and its fluctuations in the polycrystalline structure [42], any defects, domain wall structure modification close to an abrupt modulation, or spin-polarized-current-induced effects. Nevertheless, the analytical model developed here is a simple scaling law which may be useful in resolving experimental and nanofabrication issues.

**ACKNOWLEDGEMENT**

We acknowledge financial support from the French National Research Agency (ANR) (Grant No. JCJC MATEMAC-3D) and from the European Union's Seventh Framework Programme (FP7/2007-2013) under grant agreement No.309589 (M3d). J. A. F.-R. is grateful for financial support from the Spanish MINECO under MAT2016-76824-C3-1-R and co-support from the ESF though BES-2014-068789 and EEBB-I-16-10934.

**REFERENCES**

[1] A. Fernandez-Pacheco, R. Streubel, O. Fruchart, R. Hertel, P. Fischer, R. P. Cowburn, Nature Comm. **8**, 15756 (2017).

[2] A. Mourachkine et al., Nano Lett. **8**, 11 (2008).

[3] X. Kou et al., Adv. Matter. 23, 1393 (2011).

[4] M. F. Contreras et al., Int. J. Nanomedicine **10**, 2141 (2015).

[5] K. Pitzschel, J. Bachmann, S. Martens, J. M. Montero-Moreno, J. Kimling, G. Meier, J. Escrig, K. Nielsch, D. Görlitz, J. Appl. Phys. **109**, 033907 (2011).

[6] S. Da Col, M. Darques, O. Fruchart, and L. Cagnon, Appl. Phys. Lett. **98**, 112501 (2011).

[7] S.S. Parkin et al., Science **320**, 190 (2008).

[8] A. Berganza, C. Bran, M. Jaafar, M. Vazquez, A. Asenjo, Sci. Rep. **6**, 29702 (2016).

[9] S. Da Col, S. Jamet, M. Stano, B. Trapp, S. Le Denmat, L. Cagnon, J. C. Toussaint, O. Fruchart, Appl. Phys. Lett. **109**, 062406 (2016).

[10] M. Yan, A.Kákay, S. Gliga, and R. Hertel. Phys. Rev. Lett. **104**, 057201 (2010).

[11] S. Moretti,V. Raposo, E. Martinez, and L. Lopez-Diaz. Phys. Rev. B **95**, 064419 (2017).

[12] E. Berganza, M. Jaafar, C. Bran, J. A. Fernández-Roldán, O. Chubykalo-Fesenko, M. Vázquez, A. Asenjo, Sci. Rep. **7**, 11576 (2017).

[13] N. Biziere, R. Lassale Ballier, M. C. Clochard, M. Viret, T. L. Wade, E. Balanzat, J. E. Wegrowe, J. Appl. Phys **110**, 063906 (2011).

[14] A. S. Esmaeily, M. Venkatesan, A. S. Razavian, J. M. D. Coey, J. Appl. Phys. **113**, 17A327 (2013).

[15] O. Iglesias-Freire, C. Bran, E. Berganza, I. Mínguez-Bacho, C. Magén, M. Vázquez, A. Asenjo, Nanotech. **26**, 395702 (2015).

[16] E. M. Palmero, C. Bran, R. P. del Real and M. Vázquez. J. Phys.: Conf. Ser. **755**, 012001 (2016).

[17] V. Mohanan P and P.S. A. Kumar. J. Magn. Magn. Mater., **422**, 419 (2017).

[18] P. Bruno, Phys. Rev. Lett. **83**, 2425 (1999).

[19] S. Allende, D. Altbir, K. Nielsch, Phys. Rev. B **80**, 174402 (2009).

[20] S. Allende and R. Arias. Phys. Rev. B **83**, 174452 (2011).

[21] M. Franchin, A. Knittel, M. Albert, D. Chernyshenko, T. Fischbacher, A. Prabhakar, H. Fangohr, Phys. Rev. B **84**, 094409 (2011).

[22] M.Sturma, J.-C. Toussaint, D. Gusakova, J. Appl. Phys. **117**, 243901 (2015).

[23] L. C. C. Arzuza, R. Lopez-Ruiz, D. Salazar-Aravena, M. Knobel, F. Beron, K. R. Pirota, JMMM **432**, 309 (2017).

[24] F. Alouge, E. Kritsikis, J. Steiner, and J.-C. Toussaint, Numer. Math. **128**, 407 (2014).

[25] A. Hubert and R. Schäfer, Magnetic Domains, Springer Verlag, Berlin (1998).

[26] A. Thiaville, and Y. Nakatani, Nanomagnetism and Spintronics, Elsevier (2009).

[27] A. A. Thiele, Phys. Rev. Lett. **30**, 230 (1973).

[28] The so-called dynamical Thiele's domain wall width definition is obtained for the stationary domain wall displacement without changing its profile along the invariable cross-section wire axis. It reads $\Delta_T = 2S/\int(\partial \mathbf{m}/\partial z)^2$, where $S$ is the cross-section surface and the expression is integrated over the whole volume.

[29] R. Hertel, J. Magn. Magn. Mater. **249**, 251 (2002).

[30] S. Jamet, N. Rougemaille, J.-C. Toussaint, O. Fuchart, in Magnetic Nano- and Microwires : Design, synthesis, properties and applications, M. Vázquez Ed., Woodhead (2015).

[31] A. Thiaville and Y. Nakatani, Dynamics in Confined Magnetic Structures III, Topics Applied Physics **106**, 161 (2006).

[32] Y. Nakatani, A. Thiaville, J. Miltat, Nat. Mat. **2**, 521 (2003).

[33] E. Martines, L. Lopez-Diaz, L. Torres, C. Tristan, and O. Alejos, Phys. Rev. B **75**, 174409 (2007).

[34] A. Mougin, M. Cormier, J. P. Adam, P. Metaxas, and J. Ferré, Europhys. Lett. **78**, 57007 (2007).

[35] B. Krüger, J. Phys.: Condens. Matter **24**, 024209 (2012).

[36] C. A. Ferguson, D. A. MacLaren, S. McVitie, J. Magn. Magn. Mat. **381**, 457 (2015).

[37] R. Hertel, A. Kakay, J. Magn. Magn. Mat. **379**, 45 (2015).

[38] E. Durand, Electrostatique et Magnétostatique, Masson et Cie, Paris (1953).

[39] The parameter $\delta = \lambda/4$ in $\tanh(z/\delta)$ is estimated by comparing the slopes of the circle-based wire profile used in numerical simulations and of the analytical tanh-based profile close to the modulation center. This relates $\lambda$ and $\delta$ via the expression $4\delta = \lambda(1 - (R_2 - R_1)^2/\lambda^2)$. For the gently sloping modulation with $(R_2 - R_1) \ll \lambda$ this expression reduces to $4\delta \cong \lambda$. The correction to this expression is quadratic in $(R_2 - R_1)/\lambda$. For $(R_2 - R_1)/\lambda < 0.316$ the relative error made by tahn-based shape approximation is less than 10%, which is suitable for a large range of modulation sizes. In the extreme cases of very abrupt diameter transition the disagreement between circle-based and tanh-based modulation shapes is more pronounced. Nevertheless this mismatch is largely overtopped by such model imperfection as, for example, the omission of the magnetostatic domain wall repulsion from the modulation.

[40] J. Kimling, Magnetization Reversal in Cylindrical Nanowires and in Nanowires with Perpendicular Magnetic Anisotropy. Cuvillier Verlag, Göttingen, 2013. ISBN 978-3-95404-518-1.

[41] Yu. P. Ivanov, M. Vazquez and O. Chubykalo-Fesenko. J. Phys. D: Appl. Phys. **46**, 485001 (2013).

[42] A. A. Ivanov and V. A. Orlov. Phys. Solid State **53**, 2441 (2011).

**FIGURE CAPTIONS**

FIG.1: (Color online) (a) Cylindrical nanowire with modulated diameter from smaller value $2R_1$ to larger value $2R_2$. Modulation of the length $\lambda$ is centered relative to $z=0$. (b) Circle-based layout of the modulation. (c) Tail-to-tail domain wall displacement under the applied magnetic field. The color scale bar indicates the longitudinal magnetization $m_z$.

FIG.2: (Color online) Domain wall behavior far from the modulation. (a) Schematic illustration of field-driven precessional motion of the domain wall along the $z$ axis. (b) Domain wall width as a function of the wire's radius. Open circles correspond to the tanh-profile fit. The blue cross at $R=0$ indicates the Bloch wall parameter value $\Delta_{Bloch} = \sqrt{4A_{ex}/\mu_0 M_s^2}$. Red stars correspond to the domain wall width estimated from velocities simulated using Eq.(2). (c) Steady velocity of the simulated domain wall as a function of the applied field. Solid lines correspond to values returned by Eq. (2). (d) Simulated domain wall rotational frequency in response to the field applied. The solid line corresponds to application of Eq.(3).

FIG.3: (Color online) Relationship between energy and domain wall width determined from free drift of the domain wall. The red vertical dash indicates the domain wall's initial position. (a) The internal energy of the system as a function of domain wall position. (b) Domain wall width as a function of its position. In (a) and (b) $\lambda=100$ nm and $R_1=5$ nm. Horizontal gray lines correspond to values for a straight wire. (c) Sketch of the domain wall width modification when entering or leaving the modulation. Hatched area schematizes the total volume occupied by the corresponding magnetic charge.

FIG.4: (Color online) (a) Domain wall position over time. (b) Internal energy of the system as a function of domain wall position. (c) Domain wall width depending on its position. In (a), (b) and (c) $\lambda=100$ nm, $R_2=12.5$ nm and $R_1=5$ nm.

FIG.5: (Color online) (a) Schematic illustration of the uniformly charged sphere corresponding to the domain wall. (b) Domain wall energy $E_0$ as a function of the perfect wire radius in the absence of applied field: comparison between model values and simulations. (c) Domain wall energy $E_0$ as a function of the domain wall's position for several values of applied magnetic field. (d) Magnetic field distribution along the wire axis induced by the modulation charges. (e) Domain wall position as a function of the applied field, determined by applying Eq.(8) (red dashed line) and Eq.(7) (black line). The following geometry parameters were used: $R_1=5$ nm, $R_2=10$ nm and $\lambda=100$ nm.

FIG.6: (Color online) (a) Change to the critical field value depending on the larger radius $R_2$, as determined using Eq.(7) for $\lambda=100$ nm. (b) Change to the critical field value as a function of modulation length $\lambda$, as determined by applying Eq.(7) for $R_1=5$ nm.

FIG.7: (Color online) Comparison between simulated values, Eq.(7) and Eq.(9). (a) Changes to critical field value as a function of larger radius $R_2$. All curves are plotted for $R_1=5$ nm and $\lambda=100$ nm. (b) Changes to critical field value as a function of modulation length, $\lambda$. All curves are plotted for $R_1=5$ nm and $R_2=6$ nm.

FIG.8: (Color online) Comparison between simulated values, Eq.(7) and Eq.(8). Final domain wall position depending on the field applied. (a) $R_1=5$ nm, $R_2=6$ nm, $\lambda=75$ nm; (b) $R_1=5$ nm, $R_2=6$ nm, $\lambda=100$ nm; (c) $R_1=5$ nm, $R_2=10$ nm, $\lambda=100$ nm.

FIG.9: (Color online) Isovalues of a magnetic charge $\rho$ induced by the presence of the domain wall (a) close to modulation and (b) far from the modulation for simulated domain wall ($\rho = -M_s \, div\mathbf{m}$) and uniformly charged sphere ($\rho = cts$).

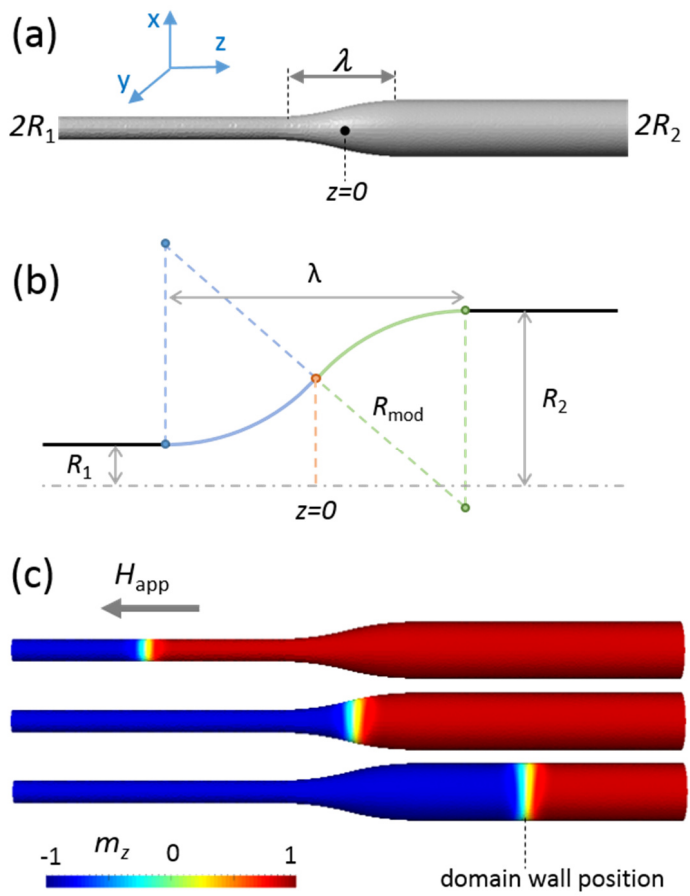

*Fernandez-Roldan et al*                                        FIG.1

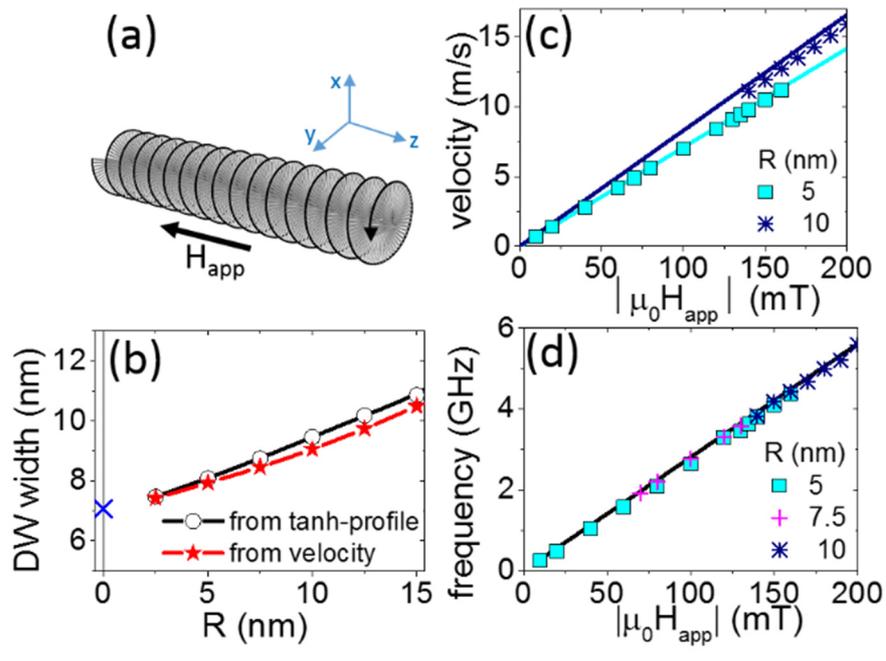

*Fernandez-Roldan et al*   FIG.2

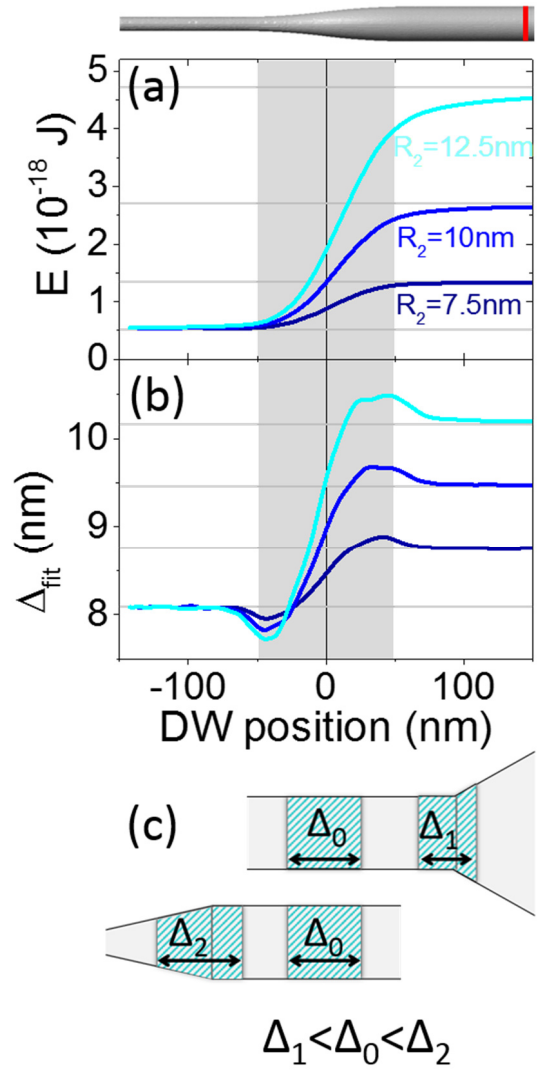

*Fernandez-Roldan et al*                                               FIG.3

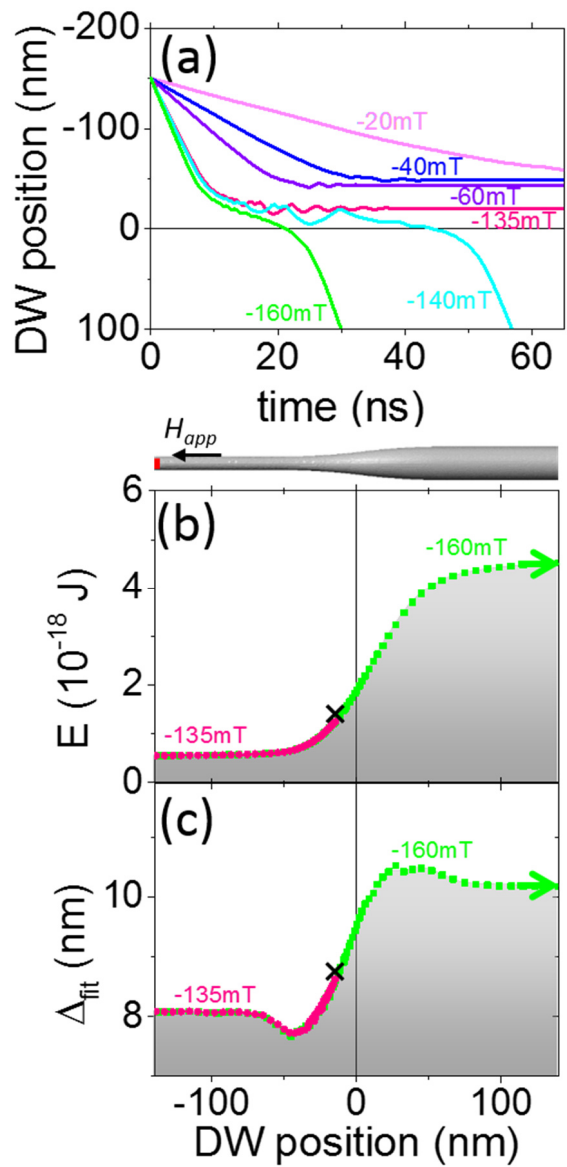

*Fernandez-Roldan et al*                                          FIG.4

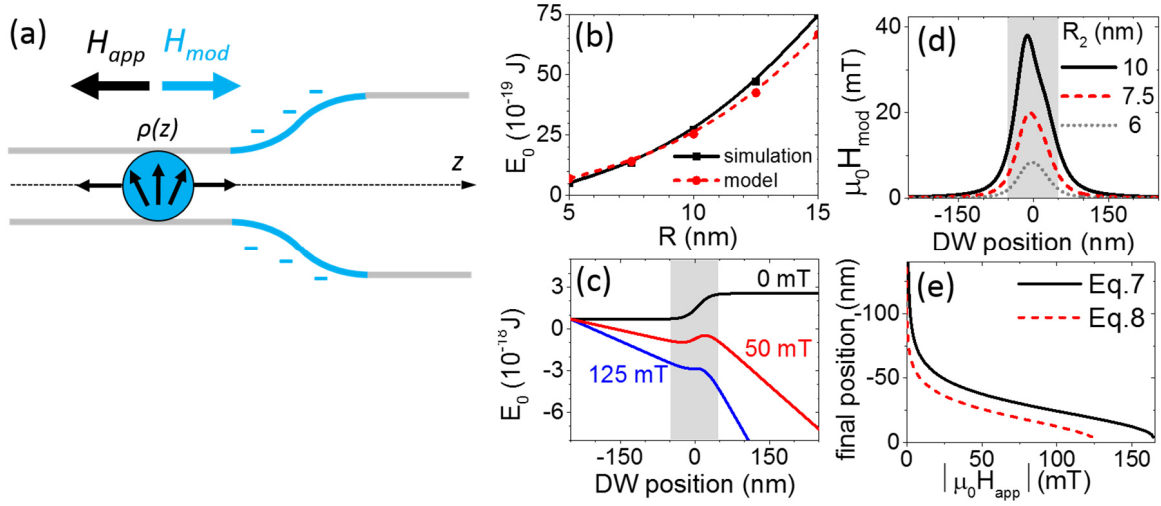

*Fernandez-Roldan et al*                          FIG.5

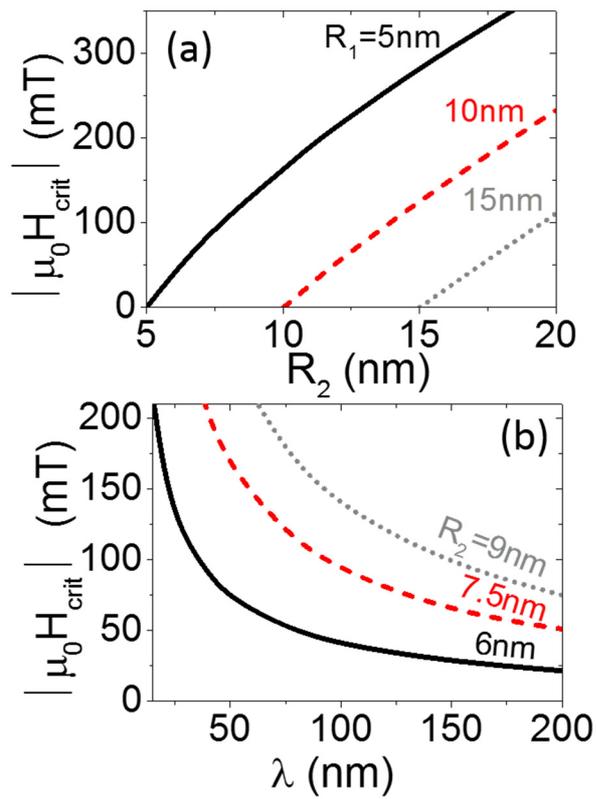

*Fernandez-Roldan et al*  FIG.6

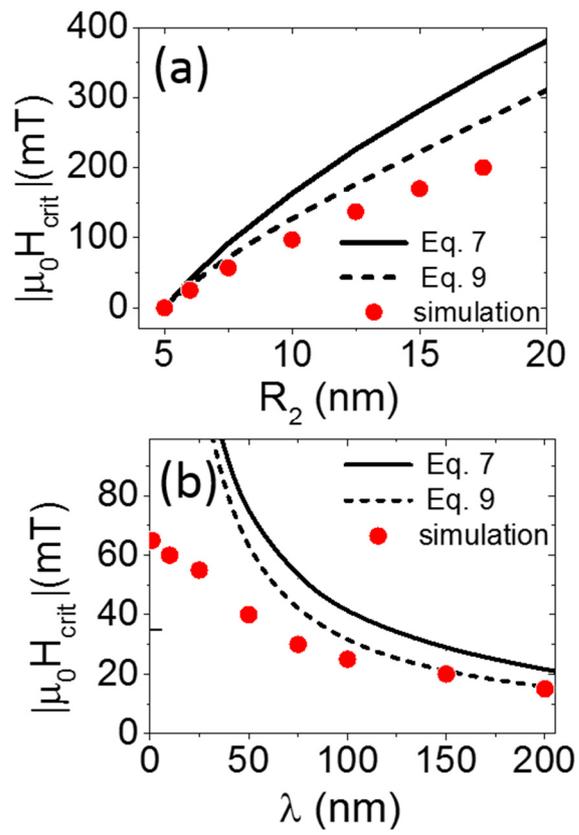

*Fernandez-Roldan et al*                                        FIG.7

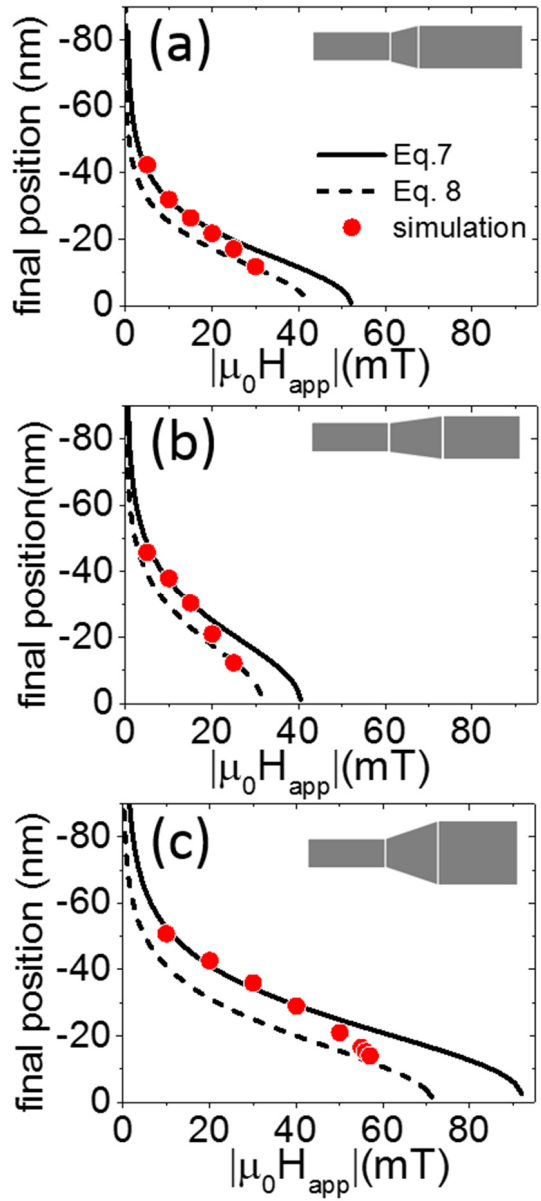

*Fernandez-Roldan et al*  FIG.8

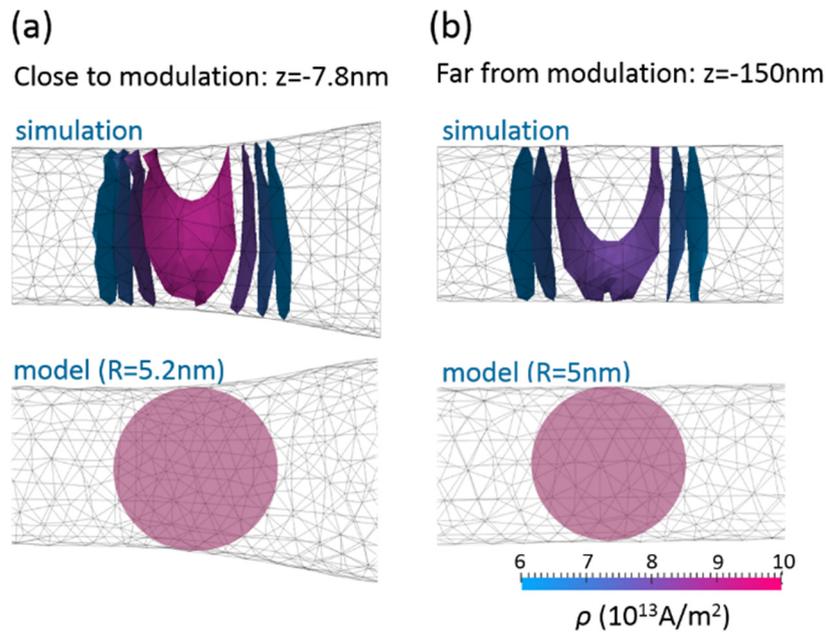

*Fernandez-Roldan et al*                                              FIG.9